\documentclass[12pt,oneside]{article}
\usepackage{amssymb,amsmath,latexsym,amsthm,amsfonts}
\usepackage[english]{babel}
\usepackage[latin1]{inputenc}
\usepackage{color,hyperref}
\usepackage{slashbox}
\usepackage{multirow}
\usepackage{graphics}
\usepackage{graphicx}
\newtheorem{thm}{Theorem}
\newtheorem{prop}{Proposition}

\newtheorem{rem}{Remark}

\numberwithin{ex}{section} \numberwithin{rem}{section}
\numberwithin{equation}{section} \numberwithin{thm}{section}
\numberwithin{lem}{section} \numberwithin{coro}{section}

\def\1g{1\hskip -3pt \mbox{l}}

\title{Estimating the predictability of economic and financial time series}

\author{
{\sc Quentin Giai Gianetto\footnote{Télécom Bretagne, 655 avenue du technopôle - 29280 Plouzané, France, quentin2g@yahoo.fr}, Jean-Marc Le Caillec\footnote{Télécom Bretagne, 655 avenue du technopôle - 29280 Plouzané, France, jm.lecaillec@telecom-bretagne.eu}, Erwan Marrec\footnote{Fédéral Finance Gestion, 1 allée Louis Lichou - 29480 Le Relecq-Kerhuon, France} 
}}

\begin{document}

\maketitle  \noindent {\em Abstract:}
The predictability of a time series is determined by the sensitivity to initial conditions of its data generating process. In this paper our goal is to characterize this sensitivity from a finite sample by assuming few hypotheses on the data generating model structure. In order to measure the distance between two trajectories induced by a same noisy chaotic dynamic from two close initial conditions, a symmetric Kullback-Leiber divergence measure is used. Our approach allows to take into account the dependence of the residual variance on initial conditions. We show it is linked to a Fisher information matrix and we investigated its expressions in the cases of covariance-stationary processes and $ARCH(\infty)$ processes. Moreover, we propose a consistent non-parametric estimator of this sensitivity matrix in the case of conditionally heteroscedastic autoregressive nonlinear processes. Various statistical hypotheses can so be tested as for instance the hypothesis that the data generating process is ``almost" independently distributed at a given moment. Applications to simulated data and to the $S\& P 500$ index illustrate our findings. More particularly, we highlight a significant relationship between the sensitivity to initial conditions of the daily returns of the $S\&P 500$ and their volatility.

\vspace*{.2cm} \noindent {\em Keywords:} {\small Chaos theory; Sensitivity to initial conditions; Non-linear predictability; Time series.}\\



\section*{Introduction}
\label{S1}


$\quad$ Stock price dynamics are difficult to approximate because of various factors influencing the supply-demand interactions. These factors can be from a political, monetary, economic or psychological nature and are difficultly measurable in real-time. However, for an investor wishing to preserve his capital, modelling the price dynamics is a necessary task to quantify investment risks and to hedge his portfolio. In this regard, the existence of exploitable deterministic chaotic dynamics has become one of the key questions in the academic literature investigating nonlinear dynamics in financial and economic time series (see Brock (1986), Hsieh (1991), Peters (1994), Hommes (2001), Shintani and Linton (2003), Kyrtsou et al. (2004), Hommes and Manzan (2005)). 

The idea behind a chaotic data generating process is that future realizations of this process can be approximated by realizations following past realizations close to the current realizations. Forecasts of such a time series can so be performed just by weighting some selected past observations. This is due to the fact that two trajectories induced by a same nonlinear chaotic dynamic will be close, until a certain time horizon, if they are generated from two close initial conditions. In contrast, when the data generating process is independent from initial conditions as in the case of independently distributed processes, further realisations cannot be determined from past values. Measuring the sensitivity of a time series to initial conditions can so indicate if it can be predicted just by using its past values.



In the literature, numerous nonlinear parametric models have been proposed to model economic and financial time series as the GARCH models (Engle (1982), Bollerslev (1986)), the threshold models (Tong (1983)) or the hidden Markov models (Mamon and Elliott (2007)). However, when we observe real-world time series, we do not know the structure of the data generating process. Nonparametric regression techniques represent an alternative to these nonlinear parametric models assuming fewer hypotheses on the model structure.
When time series are generated from a deterministic chaotic system added by a stochastic measurement noise, these regression methods can be applied to estimate the underlying chaotic dynamic. In this framework, the chaotic component, also called the ``skeleton" (Tong (1990)), models the sensitivity of the considered system to its initial conditions until a certain time horizon while the stochastic component introduces a part of unpredictability within data. More particularly, the stochastic perturbation can display heteroscedasticity, i.e. a time-varying conditional variance, which is a common feature in economic and financial time series (see Bollerslev, Chou and Kroner (1992)). In that case the conditional expectation of the process and the conditional variance of the dynamic noise can both depend on initial conditions. That is why we were interested to estimate the dependence on initial conditions of such a noisy chaotic dynamic by using nonparametric regression techniques. 

Several method already exist to measure the dependence of a time series on initial conditions. Two widely used methods are the correlation dimension introduced by Grassberger and Procaccia (1983) and the Lyapunov exponent (see Wolf et al. (1985), Rosenstein, Collins and Deluca (1993)). Such methods have initially been created for deterministic data generating process which are not perturbed by dynamic noises (see Dämming and Mitschke (1993), Tanaka, Aihara and Taki (1998)). Nonetheless, when a stochastic noise is assumed, several studies have proposed to estimate the deterministic conditional expectation of the process and its derivatives by using some non-parametric regression tools (as local polynomial non-parametric regressions, neural networks regressions, etc.). It has to be remarked that they generally assume a constant residual variance. They next compute a correlation dimension (Kawaguchi and Yanagawa (2001), Kawaguchi et al. (2005)) or a Lyapunov exponent (McCaffrey et al. (1992), Nychka et al. (1992), Gençay (1996), Lu and Smith (1997), Shintani and Linton (2004)) from the estimated conditional expectation of the process.

Anyway, these methods were originally developed for deterministic systems that is why several studies question their estimation in a stochastic context. For instance, Schittenkopf, Dorffner and Dockner (2000) used neural networks regression to estimate the Lyapunov exponent of random dynamical systems and found difficulty interpretable results. Dennis et al. (2003) developed examples of ecological population models in which a Lyapunov exponent estimated from raw data leads to conclusions opposite to those that can be deduced with a Lyapunov exponent estimated from the deterministic conditional expectation of the process. Kyrtsou and Serletis (2006) estimated a significantly negative Lyapunov exponent from daily returns of the USD/CAD exchange rate and remark that the presence of dynamic noise makes it impossible to distinguish between noisy chaos and pure randomness. Then it seems interesting to develop others methods to analyse the dependence of a noisy chaotic system on its initial values. 

For this purpose, in this paper, we propose the use of a symmetric Kullback-Leiber divergence measure applied to two distributions having different initial conditions. This measure can be linked to a Fisher information matrix (see Yao and Tong (1994)). The charm of this method rests on the fact that it allows to take into account the dependence of the residual variance on initial conditions. Schittenkopf, Dorffner and Dockner (2000) already studied this approach and gave expressions of such a Fisher information matrix when data are generated by stationary autoregressive models. Here the Fisher information matrix is estimated with local polynomial regressions what allows to take into account some non-stationary time series. A test based on this approach is next proposed to quantify the dependence on initial conditions of the data generating process. The finite sample properties of our approach are investigated through a simulation study and an application to the $S\&P500$.


Our paper proceeds as follows. In Section 2, a measure of the divergence of two initially nearby trajectories is introduced. We show that it is linked to a Fisher information matrix. Its expression is given in case of conditionally heteroscedastic nonlinear autoregressive process and in the particular cases of covariance-stationary and $ARCH(\infty)$ processes. In Section 3, an estimation of the Fisher information matrix characterizing the dependence of the data generating process on initial conditions is presented. The asymptotic properties of this estimator are studied. In Section 4, a statistical test is proposed with the aim to test the dependence on initial conditions of a data generating process from a finite sample of its realizations. In Section 5, applications on simulated data and to the $S\&P500$ index are performed. Finally, Section 6 corresponds to our conclusion.

\section{Measuring dependences on initial conditions in a noisy chaos context}

Let $(x_{t})_{t\in[1,\ldots,T]}$ be an observed time series. According to the Taken's delay embedding theorem (Takens (1981)) and its generalisations (Sauer et al. (1991)), 
if these observations are generated from a dynamic of states following some regularity conditions (see Takens (1981) for most details), 
the dynamic of these observations is fully captured in the $d$-dimensional phase space defined by the delay vectors $X(t,\tau,d)$  :
\begin{equation}\forall t\in[(d-1)\tau+1,\ldots,T],\quad X(t,\tau,d)=(x_{t},x_{t-\tau},\ldots,x_{t-(d-1)\tau})^{Tr}
\label{eqtimedel}
\end{equation}
where $\tau\in\mathbb{N}$ is the time delay, $d\in\mathbb{N}^{*}$ is a sufficiently large embedding dimension and $X^{Tr}$ denotes the transposed vector of $X$. 
In the sequel of this paper, $X(t,\tau,d)$ will be denoted by $X_{t}$ in order to simply the notations. In this framework, we thus have $x_{t+s}=f(X_{t})$ for $s\in]0,+\infty[$ where $f$ is a deterministic function respecting some regularity conditions.
However, real observations often display a behaviour which seems generated by a mix between a totally deterministic process and a totally stochastic process. That is why we consider the following conditionally
heteroscedastic nonlinear autoregressive process in this paper :
\begin{equation}
x_{t+s}=f(X_{t})+g^{\frac{1}{2}}(X_{t})\epsilon_{t}
\label{eq1}
\end{equation}
where :\begin{itemize}
\item $f$ and $g$ belong to regular spaces of functions, $g$ being a strictly positive variance function.
\item $\epsilon_{t}$ is a random variable with an independent Gaussian distribution centered on 0 and with a variance of $1$.
\end{itemize}
$f(X_{t})$ represents a component of the signal which is sensitive to initial conditions while $g^{\frac{1}{2}}(X_{t})\epsilon_{t}$ represents a random component having its variance sensitive to the same initial conditions. In practice, the regular spaces of functions are generally specified in order to use a specific statistical method to estimate $f$ and $g$ from the observations. In the sequel of this paper, we will assume that $f$ and $g$ are four times differentiable on $\mathbb{R}^{d}$. 

\subsection{A divergence measure of two nearby trajectories}

As remarked by Yao and Tong (1994), a symmetric Kullback-Leiber divergence, also known as the $J$-divergence (see Jeffreys (1946)), can be used to quantify the divergence of two initially nearby trajectories in the framework (\ref{eq1}). It equals to 
\begin{equation}
KL(t,s,\delta)=\int_{\mathbb{R}}(\rho(x_{t+s}|X_{t}+\delta)-\rho(x_{t+s}|X_{t}))\log{\Big(}\frac{\rho(x_{t+s}|X_{t}+\delta)}{\rho(x_{t+s}|X_{t})}{\Big)}dx_{t+s}
\label{eq2}
\end{equation}
where $\rho(x_{t+s}|X_{t})$ is the probability density function of $x_{t+s}$ conditioned on the delay vector $X_{t}$, and where $\delta\in\mathbb{R}^{d}$ is a vector representing the difference between two initially nearby trajectories. More particularly, in our framework (\ref{eq1}), 
\[\displaystyle{\rho(x_{t+s}|X_{t})=\frac{1}{\sqrt{2\pi g(X_{t})}}\exp{\Big(}-\frac{(x_{t+s}-f(X_{t}))^{2}}{2 g(X_{t})}{\Big)}}\]
For a fixed $\delta$, the more the system will depend on its initial conditions, the more $KL(t,s,\delta)$ will be high. It has to be noted that $KL(t,s,\delta)$ is non-negative, symmetric in $\rho(x_{t+s}|X_{t})$ and $\rho(x_{t+s}|X_{t}+\delta)$ and equals to 0 if and only if  $\rho(x_{t+s}|X_{t})=\rho(x_{t+s}|X_{t}+\delta)$. A Taylor expansion of $\rho(x_{t+s}|X_{t}+\delta)$ up to the first order in (\ref{eq2}) allows us to write that :
\begin{equation}KL(t,s,\delta)=\delta^{Tr} I(X_{t}) \delta+o_{\delta\rightarrow \mathbf{0}_{d}}(\|\delta\|^{2})\label{eqK}\end{equation}
with $I(X(t,\tau,d))$ the Fisher information matrix defined by :
\[I(X_{t})=\int_{\mathbb{R}}\frac{1}{\rho(x_{t+s}|X_{t})}\nabla\rho(x_{t+s}|X_{t})\nabla\rho(x_{t+s}|X_{t})^{Tr}dx_{t+s}\]
where $\nabla$ denotes the gradient operator with respect to the coordinates of $X_{t}$. $I(X_{t})$ is a matrix of dimensions $d\times d$ which we shall also call a {\it sensibility} matrix. 
The next proposition extracted of Schittenkopf, Dorffner and Dockner (2000) gives the expressions of $I(X_{t})$ in our framework (\ref{eq1}) :

\begin{prop} Let us assume that $f\in C^{1}(\mathbb{R}^{d})$ and $g\in C^{1}(\mathbb{R}^{d})$. Hence
\begin{equation}I(X_{t})=\frac{1} {g(X_{t})}\nabla f(X_{t})\nabla f(X_{t})^{Tr}+\frac{1}{2g(X_{t})^{2}}\nabla g(X_{t})\nabla g(X_{t})^{Tr}\end{equation}
\label{lem1}
\end{prop}

A proof of this proposition is given in the Appendix. More particularly, in the case of a constant variance $g(X_{t})=\sigma^{2}>0$, we have 
\begin{equation}
I(X_{t})=\frac{1} {\sigma^{2}}\nabla f(X_{t})\nabla f(X_{t})^{Tr}
\label{eqinf}
\end{equation}
Let us consider various situations to interpret $I(X_{t})$ in this case :
\begin{itemize}
\item If $\sigma^{2}\rightarrow+\infty$ and $0<\|\nabla f(X_{t})\|_{2}<+\infty$, the variance of $\epsilon_{t}$ masks the sensitivity of the system to its initial conditions and the symmetric Kullback-Leiber divergence will thus tend toward 0. 
\item If $0<\sigma^{2}<+\infty$ and $\|\nabla f(X_{t})\|_{2}=0$, the symmetric Kullback-Leiber divergence equals to 0 and the system is clearly not sensitive to its initial conditions since $f(X_{t})=f(X_{t}+\delta)$ when $\delta\rightarrow\mathbf{0}_{d}$. 
\item If $\sigma^{2}\rightarrow 0$ and $0<\|\nabla f(X_{t})\|_{2}<+\infty$, the system will be totally depending on its initial conditions because $\epsilon_{t}$ will tend to 0 in probability. Consequently, $KL(t,s,\delta)$ will tend to $+\infty$.
\end{itemize}

\subsection{Dependence on initial conditions of covariance-stationary processes and $ARCH(\infty)$ processes}

In this part, we study the dependence on initial conditions of specific random processes widely used in econometrics by computing the previously introduced sensibility matrix.

\paragraph{Covariance-stationary processes.} The Wold decomposition (see Hamilton (1994), p.109) ensures that any zero-mean covariance-stationary process can be represented as 
\[x_{t}=\mu_{t}+\gamma_{t}\]
where :\begin{itemize}
\item $\gamma_{t}=\sigma\sum_{j=0}^{\infty}\phi_{j}\epsilon_{t-j}$ with $\epsilon_{t}$ a standard white noise, $\sigma>0$ a variance parameter, $\phi_{0}=1$ and $\sum_{j=0}^{\infty}\phi_{j}^{2}<+\infty$.
\item $\mu_{t}$ is a linearly deterministic component
 of $x_{t}$ which we denote by $\mu_{t}=\alpha_{0}+\sum_{j=1}^{\infty}\alpha_{j}x_{t-j}$.
\end{itemize}
By considering our previous notations, we have $X_{t}=(x_{t-1},x_{t-2},\ldots,x_{t-j},\ldots)$, $f(X_{t})=\mu_{t}$ and $g^{\frac{1}{2}}(X_{t})\epsilon_{t}=\gamma_{t}$. In that case, $f(X_{t})$ is clearly dependent on $X_{t}$ while $g(X_{t})=(\gamma_{t}/\epsilon_{t})^{2}$ is independent of $X_{t}$. So we have $\frac{\partial f(X_{t})}{\partial x_{t-j}}=\alpha_{j}$ and $\nabla g(X_{t})=0$ what give
\[I(X_{t})_{ij}=\frac{\epsilon_{t}^{2}}{\gamma_{t}^{2}}\alpha_{i}\alpha_{j}\quad\quad\forall(i,j)\in\mathbb{N}^{*,2}\]
where $I(X_{t})_{ij}$ denotes the coefficient on the $i^{\text{th}}$ row and $j^{\text{th}}$ column of the
matrix $I(X_{t})$.
The sensitivity to initial conditions is thus determined by the level of the ratio $(\gamma_{t}/\epsilon_{t})^{2}$  but also by the products of the parameters $\alpha_{i}$. 
\begin{rem}
Any ARMA(p,q) processes of the form
\[(1-\sum_{i=1}^{p}\alpha_{i}L^{i})x_{t}=\alpha_{0}+(1+\sum_{j=1}^{q}\phi_{j}L^{j})\sigma\epsilon_{t}\]
where $L$ is the lag operator, $\epsilon_{t}$ is a standard white noise and where ($\alpha_{i}$) and ($\phi_{j}$) are real parameters,
has a $p\times p$ sensibility matrix equals to
\[I(X_{t})_{ij}=\frac{\epsilon_{t}^{2}}{\sigma^{2}(\epsilon_{t}+\sum_{j=1}^{q}\phi_{j}\epsilon_{t-j})^{2}}\alpha_{i}\alpha_{j}\quad\quad\forall(i,j)\in[1,\ldots,p]^{2}\]
In the specific case of an autoregressive process of order $p$, we have \[I(X_{t})_{ij}=\frac{1}{\sigma^{2}}\alpha_{i}\alpha_{j}\quad\quad\forall(i,j)\in[1,\ldots,p]^{2}\]
what means that the sensitivity to initial conditions of any autoregressive process is unchanging over time.
\end{rem}

\paragraph{ARCH($\infty$) processes.} Now, we consider an ARCH($\infty$) process which can be noted by
\[x_{t}=\sigma_{t}\epsilon_{t}\]
where $\epsilon_{t}$ is a standard white noise and 
\[\sigma^{2}_{t}=\omega_{0}+\sum_{j=1}^{\infty}\omega_{j}x_{t-j}^{2}\]
where $\omega_{j}>0$ and $\sum_{j=1}^{\infty}\omega_{j}<+\infty$. Let us recall that ARCH($m$) and GARCH($p,q$) processes can be considered as particular cases of such a process (see Hamilton (1994), p.665).
Here, we have $f(X_{t})=0$ and $g(X_{t})=\sigma^{2}_{t}$ what gives $\frac{\partial g(X_{t})}{\partial x_{t-j}}=2\omega_{j}x_{t-j}$. Hence,
\[I(X_{t})_{ij}=\frac{2\omega_{i}\omega_{j}x_{t-i}x_{t-j}}{(\omega_{0}+\sum_{j=1}^{\infty}\omega_{j}x_{t-j}^{2})^{2}}\quad\forall(i,j)\in\mathbb{N}^{*,2}\]
In that case, the dependence on initial conditions is function of the parameters $\omega_{j}$ but also of the past values of the time series. This dependence is thus time-varying similarly to a moving average process and contrary to an autoregressive process. 

\

When we work with real time series, a major issue consists of estimating $I(X_{t})$ without knowledge of the data generating process.
With this purpose, we propose, in the rest of this paper, a consistent estimator of the sensibility matrix displayed in Proposition \ref{lem1} by using local polynomial non-parametric regressions. This estimator will allow to measure the dependence on initial conditions of any observed time series having a dynamic respecting our framework displayed in \ref{eq1}. 

\section{Local estimation of the sensibility matrix}

The Fisher information matrix displayed in Proposition \ref{lem1} depends on $\nabla f(X_{t})$, $\nabla g(X_{t})$ and $g(X_{t})$. In order to estimate these quantities, we propose to use local polynomial non-parametric regression which is a widely used method displaying various advantages (see Fan and Gijbels (1996) for most details).

\subsection{Estimation by local polynomial regression}

This method begins with the following two steps :
\begin{itemize}
\item $\textbf{Phase space reconstruction of the observed time series }x_{t}$. This steps consists of estimating the time delay $\tau$ and the minimal embedding dimension $d_{min}$. Some references can be made to Fraser and Swinney (1986) or Moon, Rajagopalan and Lall (1995) concerning the estimation of $\tau$. Several algorithms are available for the estimation of $d_{min}$ (see for instance Fraser and Swinney (1986), Kennel, Brown and Abarbanel (1992), Cao (1997), Kantz and Shreiber (2003)). Popular methods are the False Nearest neighbors method (Fraser and Swinney (1986)) or the Cao method (Cao (1997)). When $\tau$ and $d_{min}$ are adequately chosen, the delay vectors can be reconstructed as in (\ref{eqtimedel}) by taking $d>d_{min}$.

\item $\textbf{Searching the }k\textbf{ nearest neighbors of each delay vector }X_{t}$ by using an Euclidean distance. If you compare each delay vector to each others, this method needs $O(T^{2})$ operations. The number of operations can be reduced to $O(T\log(T))$ if you use the k-d. tree method (see Bentley (1975), Friedman, Bentley and Finkel (1977)). The C++ library ANN allows to use such algorithms (see Arya et al. (1998)). In the sequel, we will write $t_{i}$ for the instant of the $i^{th}$ nearest neighbor of $X_{t}$. 
\end{itemize}

\paragraph{Estimation of $f(X_{t})$ and $\nabla f(X_{t})$.} Let $X_{t_{i}}$ be in a neighborhood of $X_{t}$. $f(X_{t_{i}})$ can thus be approximated by the Taylor series expansion up to the second order given by :
\begin{equation}
f(X_{t_{i}})\approx f(X_{t})+(X_{t_{i}}-X_{t})^{Tr}\nabla f(X_{t})+\frac{1}{2}(X_{t_{i}}-X_{t})^{Tr}\nabla^{2} f(X_{t})(X_{t_{i}}-X_{t})\label{eqtay}\end{equation}
where $\nabla^{2} f(X_{t})$ is the Hessian matrix of $f$ with respect to the coordinates of $X_{t}$.
The local non-parametric polynomial regression is based on this approach and consists of minimising the locally weighted sum of squared residuals :
\begin{equation}\widehat{\mathbf{B}}_{f,H}(t)=\arg\min_{\mathbf{B}}\begin{pmatrix}\mathbf{P}_{s}(t)-\mathbf{D}(t)\mathbf{B}\end{pmatrix}^{Tr}\mathbf{K}_{H}(t)\begin{pmatrix}\mathbf{P}_{s}(t)-\mathbf{D}(t)\mathbf{B}\end{pmatrix}
\label{eqmini}
\end{equation}
where :
\begin{itemize}
\item $\mathbf{D}(t)=\begin{pmatrix}1&(X_{t_{1}}-X_{t})^{Tr}&\textbf{vech}((X_{t_{1}}-X_{t})(X_{t_{1}}-X_{t})^{Tr})^{Tr}\\
\vdots&\vdots&\vdots\\
1&(X_{t_{k}}-X_{t})^{Tr}&\textbf{vech}((X_{t_{k}}-X_{t})(X_{t_{k}}-X_{t})^{Tr})^{Tr}\end{pmatrix}$

where $\textbf{vech}(A)$ denotes a vector containing the columns on and below the diagonal of a matrix $A$. $(X_{t_{i}})_{i\in[1,\ldots,k]}$ represents the $k$ nearest delay vectors of $X_{t}$ in the sense of an Euclidean distance.
\item $\mathbf{P}_{s}(t)=\begin{pmatrix}x_{t_{1}+s},\ldots,x_{t_{k}+s}\end{pmatrix}^{Tr}$ contains the realizations following the $k$ nearest neighbors of the delay vector $X_{t}$.
\item $\mathbf{K}_{H}(t)$ is a weighing matrix such that \begin{equation}\mathbf{K}_{H}(t)=\begin{pmatrix}K_{H}(X_{t_{1}}-X_{t})&\ldots &K_{H}(X_{t_{k}}-X_{t})\end{pmatrix}^{Tr}\mathbf{I}_{k}\label{kern}\end{equation} where $K_{H}(X)=\frac{1}{det(H)}K(H^{-1}X)$ with $K$ a kernel function and $\mathbf{I}_{k}$ is an identity matrix of dimension $k\times k$. For simplicity we consider $K$ spherically symmetric and $H=h\mathbf{I}_{d}$ where $h\in\mathbb{R}$ in the sequel of this paper. A common choice for $K$ is the standard normal density function $K(X)=(2\pi)^{-\frac{d}{2}}e^{-\frac{\|X\|^{2}}{2}}$.
\end{itemize}

The first derivative of (\ref{eqmini}) with respect to $\mathbf{B}$ allows to find that
\begin{equation}\widehat{\mathbf{B}}_{f,H}(t)=(\mathbf{D}(t)^{Tr}\mathbf{K}_{H}(t)\mathbf{D}(t))^{-1}\mathbf{D}(t)^{Tr}\mathbf{K}_{H}(t)\mathbf{P}_{s}(t)
\label{estim}
\end{equation}
where $\widehat{\mathbf{B}}_{f,H}(t)$ is an estimation of $\displaystyle{\mathbf{B}_{f}(t)=\begin{pmatrix}f(X_{t})&\nabla f(X_{t})^{Tr}& (\textbf{vech}(L))^{Tr}\end{pmatrix}^{Tr}}$ with $L_{ij}=\nabla^{2}f(X_{t})_{ij}$ if $i\neq j$ and $L_{ij}=\frac{1}{2}\nabla^{2}f(X_{t})_{ij}$ if $i=j$. 
The $1^{st}$ coordinate of $\widehat{\mathbf{B}}_{f,H}(t)$ corresponds to an estimation of $f(X_{t})$, denoted $\widehat{f}_{h}(X_{t})$ in the case $H=h\mathbf{I}_{d}$, while the $2$ to $d+1$ coordinates correspond to an estimation of $\nabla f(X_{t})$ denoted $\widehat{\nabla f}_{h}(X_{t})$. 

\paragraph{Estimation of $g(X_{t})$ and $\nabla g(X_{t})$.}  Now let us denote the residual vector $\widehat{\delta}(t)=\mathbf{P}_{s}(t)-\mathbf{D}(t)\widehat{\mathbf{B}}_{f,H}(t)$ and $\widehat{\delta}^{2}(t)$ the vector containing the squares of the coordinates of $\widehat{\delta}(t)$. 

With the aim of estimating $g(X_{t})$ and $\nabla g(X_{t})$, we use the following local non-parametric estimator based on the residuals of the local non-parametric estimation of $\mathbf{B}_{f}(t)$ :

\begin{equation}\widehat{\mathbf{B}}_{\delta^{2},H_{2}}(t)=(\mathbf{D}(t)^{Tr}\mathbf{K}_{H_{2}}(t)\mathbf{D}(t))^{-1}\mathbf{D}(t)^{Tr}\mathbf{K}_{H_{2}}(t)\widehat{\delta}^{2}(t)
\label{eqdel}
\end{equation}

Similarly to $\widehat{\mathbf{B}}_{f,H}(t)$, the $1^{st}$ coordinate of $\widehat{\mathbf{B}}_{\delta^{2},H_{2}}(t)$ corresponds to an estimation of $\mathbb{E}[(x_{t+s}-\widehat{f}_{h}(X_{t}))^{2}]$ denoted by $\widehat{g}_{h_{2}}(X_{t})$ in the case $H_{2}=h_{2}\mathbf{I}_{d}$, while the $2$ to $d+1$ coordinates correspond to an estimation of $\nabla \mathbb{E}[(x_{t+s}-\widehat{f}_{h}(X_{t}))^{2}]$ denoted by $\widehat{\nabla g}_{h_{2}}(X_{t})$. 

\

\begin{rem}It has to be noted that the matrices $\mathbf{D}(t)^{Tr}\mathbf{K}_{H}(t)\mathbf{D}(t)$ and $\mathbf{D}(t)^{Tr}\mathbf{K}_{H_{2}}(t)\mathbf{D}(t)$ need to be invertible in (\ref{estim}) and (\ref{eqdel}). When two nearest neighbours are close to each other, their respective columns in these matrices are also close and these one will thus have determinants close to $0$. To by-pass this numerical problem, one of both close neighbours can be eliminated. An other method consists of using a regression on principal components by transforming the columns of the matrix $\mathbf{K}_{H}^{\frac{1}{2}}(t)\mathbf{D}(t)$ into a matrix with orthonormal columns (see Jolliffe (2002) for most details).
\end{rem}


\

Finally, we propose an estimator of the sensibility matrix displayed in Proposition \ref{lem1}) by using (\ref{estim}) and (\ref{eqdel}) :
\begin{equation}\widehat{I}_{h,h_{2}}(X_{t})=\frac{1}{\widehat{g}_{h_{2}}(X_{t})}\widehat{\nabla f}_{h}(X_{t})\widehat{\nabla f}_{h}(X_{t})^{Tr}+\frac{1}{2\widehat{g}_{h_{2}}(X_{t})^{2}}\widehat{\nabla} g_{h_{2}}(X_{t})\widehat{\nabla} g_{h_{2}}(X_{t})^{Tr}\label{EstimFish}\end{equation}
where $h$ and $h_{2}$ are two selected bandwidths. In the next section, we investigate the asymptotic properties of $\widehat{I}_{h,h_{2}}(X_{t})$ displayed in (\ref{EstimFish}).

\subsection{Asymptotic consistency of $\widehat{I}_{h,h_{2}}(X_{t})$}
The following theorem gives the asymptotic consistency of $\widehat{I}_{h,h_{2}}(X_{t})$ under some general conditions. 
\begin{thm} Assume that $f\in\mathcal{C}^{4}(\mathbb{R}^{d})$ and $g\in\mathcal{C}^{4}(\mathbb{R}^{d})$ in model (\ref{eq1}). Let us denote $\underset{}{\overset{P}{\longrightarrow}}$ the convergence in probability, $\mu_{l}=\int u_{1}^{l}K(u)du$ and $J_{l}=\int u_{1}^{l}K^{2}(u)du$ where $K$ is the kernel function already introduced in (\ref{kern}) and $u$ is a vector of coordinates $(u_{i})_{i\in[1,\ldots,d]}$. Assume moreover that :
\begin{itemize}
\item $0<\mu_{l}<+\infty$ and $0<J_{l}<+\infty$ for all $l\in\mathbb{N}$.
\item $\{X_{t_{i}}\}_{i\in[1,\ldots,k]}$ is a multivariate i.i.d. sequence with a marginal density noted $d_{X}$ such as $d_{X}(X_{t})>0$ and $d_{X}\in\mathcal{C}^{1}(\mathbb{R}^{d})$. 
\item $h\rightarrow 0$, $h_{2}\rightarrow 0$, $kh^{d}\rightarrow +\infty$, $kh_{2}^{d}\rightarrow +\infty$, $kh^{d+2}\rightarrow +\infty$ and $kh_{2}^{d+2}\rightarrow +\infty$ when $k\rightarrow +\infty$. 
\end{itemize}
Under these general conditions, we have 
\[\widehat{I}_{h,h_{2}}(X_{t})_{pq}\underset{k\rightarrow +\infty}{\overset{P}{\longrightarrow}}I(X_{t})_{pq}\]
where $\widehat{I}_{h,h_{2}}(X_{t})_{pq}$ and $I(X_{t})_{pq}$ denote the coefficients on the $p^{th}$ row and $q^{th}$ column of the matrix $\widehat{I}_{h,h_{2}}(X_{t})$ and $I(X_{t})$ respectively.
\label{Thm1}
\end{thm}

A proof of Theorem \ref{Thm1} is given in the Appendix. It has to be noted that the assumption of i.i.d. nearest neighbors can be relaxed as in Lu (1999) (Condition C). In the next section, we propose to use a bootstrapping technique to test the local dependence on initial conditions for a finite-length time series.

\section{Testing the local sensitivity to initial conditions within time series}
\label{test}

The dynamics following two close delay vectors are considered similar if the symmetric Kullback-Leiber divergence presented in (\ref{eq2}) between the densities $\rho(x_{t+s}|X_{t_{i}})$ and $\rho(x_{t+s}|X_{t_{j}})$ is close to 0 for $i\in[1,\ldots,k]$ and $j\in[1,\ldots,k]$, where $X_{t_{i}}$ and $X_{t_{j}}$ are two close delay vectors in the sense of the Euclidean distance. Furthermore, let us recall from (\ref{eqK}) that $KL(t,s,\delta)\approx\delta^{Tr}I(X_{t})\delta$ when $\|\delta\|^{2}$ is quite small. 

That is why we consider the following statistic :
\[S_{h,h_{2}}(t)= \exp(-U_{h,h_{2}}(t))\]
where
\[U_{h,h_{2}}(t)=\sum_{q=1}^{d}\sum_{p=1}^{d}|\widehat{I}_{h,h_{2}}(X_{t})_{pq}|\]
with $h$ and $h_{2}$ two selected bandwidths and $d$ a selected embedding dimension.
Under the general conditions of Theorem $\ref{Thm1}$ and from the continuous mapping theorem, we so have
\[S_{h,h_{2}}(t)\underset{k\rightarrow +\infty}{\overset{P}{\longrightarrow}}\exp(-\sum_{q=1}^{d}\sum_{p=1}^{d}|I(X_{t})|):=S(t)\]
Under the hypothesis that the time series is independent from initial conditions, i.e.  $I(X_{t})_{ij}=0$ for all $(i,j)\in[1,\ldots,d]^{2}$, $S(t)$ equals to 1. More the time series will be dependent on initial conditions, more $S(t)$ will be low. If the process is totally deterministic, we have $S(t)=0$. In order to test the dependence of a time series on initial conditions, the following pair of hypotheses can so be tested :
\[H_{0}:``S(t)\leq\beta"\quad\text{vs.}\quad H_{1}:``S(t)>\beta"\]
where $\beta$ corresponds to a level of dependence on initial conditions. A p-value of this test can be obtained by estimating the probability $\mathbb{P}(S_{h,h_{2}}(t)\leq \beta)$. When $\beta$ is close to 1, this p-value corresponds to a probability that the data generating process is not independent from initial conditions. On the other hand, when $\beta$ is close to 0, this p-value corresponds to a probability that the data generating process is strongly dependent on initial conditions (almost totally deterministic).

However, the asymptotic distribution of $S_{h,h_{2}}(t)$ depends on unknown quantities as the gradients $\nabla f(X_{t})$ and $\nabla g(X_{t})$ (see the proof of Theorem \ref{Thm1}). Thus $S_{h,h_{2}}(t)$ cannot directly be used to build a test and we consider a re-sampling procedure to provide reliable quantiles for testing the local sensitivity of a time series to initial conditions. This re-sampling method is inspired from Gençay (1996) who applied a similar method to approximate the distribution of a maximal Lyapunov exponent. 

Firtsly $l$ delay vectors are selected with replacement from all the $k$ selected nearest neighbors $\{X_{t_{i}}\}_{i\in[1,\ldots,k]}$ of $X_{t}$. The new selected set, denoted by $\{X_{t_{i}}^{*}\}_{i\in[1,\ldots,l]}$, allows to estimate a new $\widehat{\mathbf{B}}_{f,H}^{*}(t)$. This estimation allows to find a new $\widehat{\mathbf{B}}_{\delta^{2},H_{2}}^{*}(t)$ and, finally, a new estimation of the Fisher information matrix $\widehat{I}_{h,h_{2}}(X_{t})$ and thus of $S_{h,h_{2}}(t)$, denoted respectively by $\widehat{I}_{h,h_{2}}^{*}(X_{t})$ and $S_{h,h_{2}}^{*}(t)$, are obtained. 
When this experiment is repeated a large number of time, it allows to approximate the distribution of $S_{h,h_{2}}(t)$. In practice, it has to be noted that the $k$ nearest neighbors of $X_{t}$ needs to be significantly close to $X_{t}$ while the number $l$ of selected delay vectors needs to be choose not too small to obtain significant approximations of the sensibility matrix.


\section{Simulations and empirical applications}

In this section, we firstly applied our approach to stationary AR(1) processes displaying a constant dependence on initial conditions and to a GARCH(1,1) process where this dependence is time-varying. We next test this dependence on the daily returns of the $S\& P 500$ index. In all our experiments, we fixed the prediction horizon $s=1$.

\subsection{Results on AR(1) processes.}
Firstly, let us consider the following AR(1) process :
\begin{equation}x_{t,\phi_{1}}=0.5+\phi_{1}x_{t-1}+\sigma\epsilon_{t}\end{equation}
where $\epsilon_{t}$ is a standard white noise. The Fisher information matrix measuring the sensitivity to initial conditions of this process is given by
\[I(X_{t})=\frac{\phi_{1}^{2}}{\sigma^{2}}\]
For our experiments, we fix the variance parameter $\sigma=0.1$ and we consider three stationary AR(1) processes corresponding to the cases $\phi_{1}=0.01$, $\phi_{1}=0.5$ and $\phi_{1}=0.95$. For each stationary AR(1) process, 499 time series of size $T=1000$ are generated. The values of their respective theoretical statistical index $S(T)$ will thus be $e^{-0.01}$, $e^{-25}$ and $e^{-90.25}$. In our experiments, we assume that the time delay $\tau=1$ and the embedding dimension $d=2$. Estimations of the proposed statistic $S_{h,h_{2}}(T)$ are done by using the standard normal density kernel function in the local polynomial non-parametric regressions and by fixing $h=h_{2}=0.1$. In our re-sampling procedure (see section \ref{test}), we estimated the probabilities $\widehat{\mathbb{P}}(S_{h,h_{2}}(T)\leq 0.1)$ and $\widehat{\mathbb{P}}(S_{h,h_{2}}(T)\leq 0.9)$ from 199 iterations by choosing $l=20$ delay vectors with replacement from $k=30$ nearest neighbours of the current delay vector $X_{T}$. The empirical cumulative distribution functions of these estimated probabilities, denoted by $\widehat{\mathbb{P}}_{\phi_{1}}$, are assessed by means of Monte Carlo experiments from the 499 generated time series. Figure \ref{F1} and Figure \ref{F2} display these distribution functions for the three stationary AR(1) processes. 

In view of Figure \ref{F1} and Figure \ref{F2}, we observe that \[\widehat{\mathbb{P}}_{\phi_{1}=0.01}(\widehat{\mathbb{P}}{\big(}S_{h,h_{2}}(T)\leq\beta)\leq\alpha{\big)}\geq\widehat{\mathbb{P}}_{\phi_{1}=0.5}{\big(}\widehat{\mathbb{P}}(S_{h,h_{2}}(T)\leq\beta)\leq\alpha{\big)}\geq\widehat{\mathbb{P}}_{\phi_{1}=0.95}{\big(}\widehat{\mathbb{P}}(S_{h,h_{2}}(T)\leq\beta)\leq\alpha{\big)}\]
for all $\alpha\in[0,1]$ and $\beta\in\{0.1,0.9\}$. This illustrates the lower dependence on initial conditions for $\phi_{1}=0.01$ ($S(t)=e^{-0.01}$) than in the case $\phi_{1}=0.5$ ($S(t)=e^{-25}$) which has itself a lower dependence on initial conditions than in the case $\phi_{1}=0.95$ ($S(t)=e^{-90.25}$). 

It has to be noted that in the cases $\phi_{1}=0.5$ and $\phi_{1}=0.95$, the theoretical statistical index $S(t)$ is close to 0. However, Figure \ref{F2} clearly shows that the probability $\widehat{\mathbb{P}}_{\phi_{1}}(\widehat{\mathbb{P}}{\big(}S_{h,h_{2}}(T)\leq 0.1)=0{\big)}$ is approximatively equal to 0 for $\phi_{1}=0.95$ while it is close to 0.6 for $\phi_{1}=0.5$. The finite sample estimation of $S_{h,h_{2}}(t)$ is thus rather different from the theoretical statistical index $S(t)$ in the case of $\phi_{1}=0.5$. This difference can be due to the chosen bandwidths. Indeed, inappropriate hyper-parametrization can cause to give importance to delay vectors far from the current delay vector $X(T)$ in our non-parametric regressions and finally to conclude spuriously on the independence from initial conditions. 

Figure \ref{F3} illustrates how the empirical cumulative distribution function of $\widehat{\mathbb{P}}{\big(}S_{h,h_{2}}(T)\leq 0.9)$ can be changed by making vary the bandwidth $h$ when the others parameters are fixed ($k=30$, $l=20$ and $h_{2}=0.1$) in the case of $\phi_{1}=0.5$. Figure \ref{F4} illustrates the same thing for the bandwidth $h_{2}$ ($k=30$, $l=20$ and $h=0.1$). In view of Figure \ref{F3}, increasing the bandwidth $h$ allows to conclude on higher dependence on initial conditions of the time series. This can be explained by the fact that a too small bandwidth will give importance to very few delay vectors what can lead to spurious conclusions. However, in view of Figure \ref{F4}, changing $h_{2}$ seems to have few impact on the estimation of the empirical cumulative functions. It is logic because the AR process has not a time-varying residual variance. These observations show that the choice of hyper-parameters in the non-parametric regressions must be carefully made. More specifically, final conclusions are totally function of the chosen hyper-parametrization.

\

- Figure \ref{F1} around here -

\

- Figure \ref{F2} around here -

\

- Figure \ref{F3} around here -

\

- Figure \ref{F4} around here -

\

\subsection{Results on GARCH(1,1) process.}
Let us consider the following GARCH(1,1) process :
\begin{equation}x_{t} = \sigma_{t}\epsilon_{t}\label{eqgarch}\end{equation}
\[\sigma^{2}_{t} = \alpha_{0} + \alpha_{1}x_{t-1}^{2} + \beta_{1}\sigma^{2}_{t-1}\]
with $\alpha_{0} = 5\times 10e-6$, $\alpha_{1} = 0.05$, $\beta_{1} = 0.9$ and $\epsilon_{t}$ a standard white noise.
This process is close to those that can be infer from daily returns of stock market indices. By using an inductive reasoning and since $\|\beta_{1}\|<1$, this process can be re-written as
\[\sigma^{2}_{t} = \frac{\alpha_{0}}{1-\beta_{1}} + \alpha_{1}\sum_{i=1}^{+\infty}\beta_{1}^{i}x_{t-i}^{2}\]
Consequently, a coefficient of the Fisher information matrix is given by
\begin{equation}I(X_{t})_{ij}=\frac{2\alpha_{1}^{2}\beta_{1}^{i+j}x_{t-i}x_ {t-j}}{(\frac{\alpha_{0}}{1-\beta_{1}} + \alpha_{1}\sum_{i=1}^{+\infty}\beta_{1}^{i}x_{t-i}^{2})^{2}}\label{coefFish}\end{equation}
In order to illustrate the time-varying dependence on initial conditions of such a process, a time series of size 2000 following \ref{eqgarch} is generated. We test an hypothesis $H_{0}$ that the time series is ``almost" independently distributed by assessing the p-value $\mathbb{P}(S_{h,h_{2}}(t)\leq 1-\epsilon)$ where $\epsilon$ is close to 0. In our experiment, we fix $\epsilon=5\times 10^{-5}$. The theoretical statistical index $S(t)$ and the estimated probability $\widehat{\mathbb{P}}(S_{h,h_{2}}(t)\leq 1-\epsilon)$ are next computed from \ref{coefFish} for $t\in[1000,2000]$ by using a sliding windows of size 1000. The probability $\widehat{\mathbb{P}}(S_{h,h_{2}}(t)\leq 1-\epsilon)$ is estimated following the previous method used in the case of the AR(1) processes ($k=30$, $l=20$, $h=h_{2}=0.1$). If the probability $\widehat{\mathbb{P}}(S_{h,h_{2}}(t)\leq 1-\epsilon)=0$, the time series can be considered similar to an ``almost" independently distributed process.  Figure \ref{garch22} display the obtained outcomes. When the theoretical statistical index $S(t)$ deviates from 1, we remark that the estimated p-value increases what indicates that our method allows to get well the moments when the series is more predictable. Although $S(t)$ is never equal to 1, the estimated p-value is often close to 0 what means that the time series can often be considered as unpredictable from a past window of size 1000 with our hyper-parametrization.

\

- Figure \ref{garch22} around here -

\

\subsection{Empirical applications to the $S\&P500$}

In this section, we studied if the daily returns of the $S\&P 500$ stock market index are sensitive to initial conditions by using our approach. The considered time series goes from the 04/01/1999 to the 18/02/2010 and has a size equals to 3051. It has been extracted from Datastream. 

In order to investigate the predictability of the time series from a past data window of size 1000, we consider a sliding windows of size 1000 and apply our method with the same hyper-parameters used in the previous experiments ($k=30, l=20, h=h_{2}=0.1$). The p-value $\widehat{\mathbb{P}}(S_{h,h_{2}}(t)\leq 1-\epsilon)$ is determined with $\epsilon=1\times 10^{-4}$ and $1\times 10^{-3}$. Results are displayed in Figure \ref{sp}. When $\epsilon=1\times 10^{-4}$, the p-value varies more between 0 and 1 than in the case $\epsilon=1\times 10^{-3}$ where it is often equal to 0. The time series can so be considered predictable by using $\epsilon=1\times 10^{-4}$ while it is considered unpredictable with $\epsilon=1\times 10^{-3}$. 

The dependence on initial conditions is thus low but significant at certain moments. More particularly, we remark that the p-value is higher when the volatility of daily returns is higher what means that it is more sensitive to initial conditions in period of high volatility. Conversely, it is lower sensitive in period of low volatility. 
Figure \ref{PV} illustrates these observations by representing the probability $\widehat{\mathbb{P}}(S_{h,h_{2}}(t)\leq 1-1\times 10^{-4})$ in function of the squared daily returns of the $S\&P 500$. The Pearson's correlation coefficient between these quantities is approximatively equals to 0.3 and significantly not null. More significant relationships are established by examining the Pearson's correlation coefficients between their respective 5-day moving averages and 20-day moving averages (see Figure \ref{PV}). In view of Figure \ref{PV}, these relations are nonlinear. It has to be noted that these observations are similar to these done by LeBaron (1992) who showed that there are significant relations between volatility and serial correlations in stock market returns, serial correlations being a manner to measure the dependence on past conditions.

\

- Figure \ref{sp} around here -

\

- Figure \ref{PV} around here -

\

\section{Conclusion}

In this paper we studied the problem of testing the local sensitivity to initial conditions of time series. Our approach consists to measure the distance between two trajectories, having different initial conditions and following a same noisy chaotic dynamic, with a symmetric Kullback-Leiber divergence. We showed that this divergence can be characterized by a Fisher information matrix. In this way, we showed that autoregressive processes have a constant dependence on initial conditions while moving average processes or $ARCH(\infty)$ processes have a time-varying dependence on initial conditions. Because real-world time series have unkown data generating processes, we proposed a framework for testing the time-varying sensitivity to initial conditions of any conditionally heteroscedastic nonlinear autoregressive processes by using nonparametric regression techniques. More particularly, we propose a consistent estimator of the Fisher information matrix characterizing the dependence on initial conditions. We illustrated these theoretical results through a set of numerical experiments. We have remarked that the choice of hyper-parameters in the non-parametric regressions must be carefully made. The outcomes obtained on the daily returns of the $S\&P500$ index show that they are more sensitive to initial conditions in period of high volatility than in period of low volatility. Interesting further researches could be done by investigating the dependence on initial conditions of others time series with our method.

\newpage
\section{Appendix}
$\quad\textbf{Proof of Proposition \ref{lem1}.}$
{\small We have 
\[\nabla \log(\rho(x_{t+s}|X_{t}))=(\frac{(x_{t+s}-f(X_{t}))^{2}}{2 g(X_{t})^{2}}-\frac{1}{2 g(X_{t})})\nabla g(X_{t})\]
\[+\frac{(x_{t+s}-f(X_{t})) } {g(X_{t})}\nabla f(X_{t})\]
Hence,
\[I(X_{t})=\int_{\mathbb{R}}\rho(x_{t+s}|X_{t})\nabla \log(\rho(x_{t+s}|X_{t}))\nabla \log(\rho(x_{t+s}|X_{t}))^{Tr}dx_{t+s} \]
\[=\mathbb{E}_{x_{t+s}|X_{t}}[\frac{(x_{t+s}-f(X_{t}))^{2}} {g(X_{t})^{2}}\nabla f(X_{t})( \nabla f(X_{t}))^{Tr}\]
\[ +(\frac{(x_{t+s}-f(X_{t}))^{4}}{4 g(X_{t})^{4}}+\frac{1}{4 g(X_{t})^{2}}-\frac{(x_{t+s}-f(X_{t}))^{2}}{2 g(X_{t})^{3}})(\nabla g(X_{t}))(\nabla g(X_{t}))^{Tr}]\]
Because $x_{t+s}|X_{t}$ follows a Gaussian distribution centered on $f(X_{t})$, we have $\mathbb{E}_{x_{t+s}|X_{t}}[(x_{t+s}-f(X_{t}))^{2}]=g(X_{t})$, $\mathbb{E}_{x_{t+s}|X_{t}}[(x_{t+s}-f(X_{t}))^{3}]=0$ and $\mathbb{E}_{x_{t+s}|X_{t}}[(x_{t+s}-f(X_{t}))^{4}]=3 g(X_{t})^{2}$  what give the result for $I(X_{t})$.}

\

$\textbf{Proof of Theorem \ref{Thm1}.}$
{\small
In the next, $\underset{k\rightarrow +\infty}{\overset{d}{\longrightarrow}}$ denotes the convergence in distribution and $\underset{k\rightarrow +\infty}{\overset{P}{\longrightarrow}}$ denotes the convergence in probability. Let us consider the following theorem :
\begin{thm}
Assume that $f\in\mathcal{C}^{4}(\mathbb{R}^{d})$ and $g\in\mathcal{C}^{4}(\mathbb{R}^{d})$ in model (\ref{eq1}). Let us denote $\mu_{l}=\int u_{1}^{l}K(u)du$ and $J_{l}=\int u_{1}^{l}K^{2}(u)du$ where $K$ is the kernel function already introduced in (\ref{kern}) and $u$ is a vector of coordinates $(u_{i})_{i\in[1,\ldots,d]}$. Assume moreover that :
\begin{itemize}
\item $0<\mu_{l}<+\infty$ and $0<J_{l}<+\infty$ for all $l\in\mathbb{N}$.
\item $\{X_{t_{i}}\}_{i\in[1,\ldots,k]}$ is a multivariate i.i.d. sequence with a marginal density noted $d_{X}$ such as $d_{X}(X_{t})>0$ and $d_{X}\in\mathcal{C}^{1}(\mathbb{R}^{d})$. 
\item $h\rightarrow 0$, $kh^{d}\rightarrow +\infty$, $kh^{d+2}\rightarrow +\infty$ when $k\rightarrow +\infty$. 
\end{itemize}
Under these general conditions, we have 
\[\sqrt{k}h^{\frac{d}{2}}(\widehat{f}_{h}(X_{t})-f(X_{t})-b_{1}(X_{t},h))\underset{k\rightarrow +\infty}{\overset{d}{\longrightarrow}}\mathcal{N}(0,\sigma_{1}(X_{t}))\] 
\[\sqrt{k}h^{1+\frac{d}{2}}(\widehat{\nabla f}_{h}(X_{t})-\nabla f(X_{t})-b_{2}(X_{t},h))\underset{k\rightarrow +\infty}{\overset{d}{\longrightarrow}}\mathcal{N}(0,\Sigma_{2}(X_{t}))\] 
with :\begin{itemize}
\item $\displaystyle{b_{1}(X_{t},h)=o_{h\rightarrow 0}(h^{3})}$ and $\displaystyle{b_{2}(X_{t},h)=\frac{h^{2}}{6\mu_{2}}s(X_{t})+o_{h\rightarrow 0}(h^{3})}$ where the $i^{th}$ coordinate of $s(X_{t})$ is \[\displaystyle{s_{i}(X_{t})=\mu_{4}\frac{\partial^{3} f(X_{t})}{\partial x_{i}^{3}}+3\mu_{2}^{2}\sum_{\underset{i\neq l}{l=1}}^{d}\frac{\partial^{3} f(X_{t})}{\partial x_{l}^{2}\partial x_{i}}}\]
\item $\displaystyle{\sigma_{1}(X_{t})=\frac{g(X_{t})J_{0}}{d_{X}(X_{t})}}$ and $\Sigma_{2}(X_{t})=\sigma_{2}(X_{t})\mathbf{I}_{d}$ with $\displaystyle{\sigma_{2}(X_{t})=\frac{g(X_{t})J_{2}}{\mu_{2}^{2}d_{X}(X_{t})}}$  
\end{itemize}
\label{theo1}
\end{thm}
The proof of Theorem \ref{theo1} is similar to the proofs which can be found in Masry (1996) or Lu (1999) and is thus omitted. It has to be noted that the assumption of i.i.d. nearest neighbors can be relaxed as in Masry (1996) or Lu (1999). We deduce directly from this Theorem \ref{theo1} that each coefficient of the matrix $\widehat{\nabla f}_{h}(X_{t})\widehat{\nabla f}_{h}(X_{t})^{Tr}$ converges in distribution toward the product of two normal distributions :
\[\forall (p,q)\in[1,\ldots,d]^{2},\quad\widehat{\nabla f}_{h}(X_{t})_{p}\widehat{\nabla f}_{h}(X_{t})_{q}\underset{k\rightarrow +\infty}{\overset{d}{\longrightarrow}}Z_{p}Z_{q}\] where $\widehat{\nabla f}_{h}(X_{t})_{p}\widehat{\nabla f}_{h}(X_{t})_{q}$ denotes the coefficient of the $p^{th}$ row and the $q^{th}$ column of the matrix  $\widehat{\nabla f}_{h}(X_{t})\widehat{\nabla f}_{h}(X_{t})^{Tr}$ and  where $Z_{l}$ follows a Gaussian law $\mathcal{N}(\nabla f(X_{t})_{l}+b_{2}(X_{t},h)_{l},\frac{1}{kh^{2+d}}\sigma_{2}(X_{t}))$ with $\nabla f(X_{t})_{l}$ and $b_{2}(X_{t},h)_{l}$ the $l^{th}$ coordinates of $\nabla f(X_{t})$ and $b_{2}(X_{t},h)$ respectively. More particularly, we will have the following bias and variance for this estimator 
\begin{align*}\mathbb{E}[\widehat{\nabla f}_{h}(X_{t})_{p}\widehat{\nabla f}_{h}(X_{t})_{q}]-\nabla f(X_{t})_{p}\nabla f(X_{t})_{q}&=\nabla f(X_{t})_{p}b_{2}(X_{t},h)_{q}+b_{2}(X_{t},h)_{p}\nabla f(X_{t})_{q}+b_{2}(X_{t},h)_{p}b_{2}(X_{t},h)_{q}\\
&=o_{h\rightarrow 0}(h)\end{align*}
\begin{align*}\mathbb{V}[\widehat{\nabla f}_{h}(X_{t})_{p}\widehat{\nabla f}_{h}(X_{t})_{q}]&=\frac{\sigma_{2}^{2}(X_{t})}{(kh^{2+d})^{2}}+2\frac{\sigma_{2}(X_{t})}{kh^{2+d}}(\nabla f(X_{t})_{q}+b_{2}(X_{t},h)_{q})^{2}\text{ if }p=q\\
&=0\text{ if }p\neq q
\end{align*}
By supposing that $h\rightarrow 0$ and $kh^{2+d}\rightarrow +\infty$ when $k\rightarrow +\infty$, the consistency of $\widehat{\nabla f}_{h}(X_{t})\widehat{\nabla f}_{h}(X_{t})^{Tr}$ is thus obtained. 

The asymptotic consistency of $\widehat{g}_{h_{2}}(X_{t})$ toward $\mathbb{E}[(x_{t+s}-\widehat{f}_{h}(X_{t}))^{2}]$ can also be achieved by using the general conditions of Theorem \ref{theo1} if $h_{2}\rightarrow 0$ and $kh_{2}^{d}\rightarrow +\infty$ when $k\rightarrow+\infty$. This outcomes is obtained by replacing $f$ with $\mathbb{E}[(x_{t+s}-\widehat{f}_{h}(X_{t}))^{2}]$ and $g(X_{t})$ with a constant function in the statement of Theorem \ref{theo1}. 

Because $\widehat{f}_{h}(X_{t})$ is a consistent estimator of $f$ if $h\rightarrow 0$ and $kh^{d}\rightarrow +\infty$ when $k\rightarrow +\infty$, we have $\mathbb{E}[(x_{t+s}-\widehat{f}_{h}(X_{t}))^{2}]\underset{k\rightarrow +\infty}{\overset{P}{\longrightarrow}}\mathbb{V}[x_{t+s}-f(X_{t})]=g(X_{t})$ under these conditions. Hence, \[\widehat{g}_{h_{2}}(X_{t})\underset{k\rightarrow +\infty}{\overset{P}{\longrightarrow}}g(X_{t})\] if $h\rightarrow 0$, $h_{2}\rightarrow 0$, $kh^{d}\rightarrow +\infty$ and $kh_{2}^{d}\rightarrow +\infty$.

Similarly to $\widehat{\nabla f}_{h}(X_{t})_{p}\widehat{\nabla f}_{h}(X_{t})_{q}$, the asymptotic consistency of $\widehat{\nabla g}_{h_{2}}(X_{t})_{p}\widehat{\nabla g}_{h_{2}}(X_{t})_{q}$ can also be achieved by assuming $h_{2}\rightarrow 0$ and $kh_{2}^{d+2}\rightarrow +\infty$ when $k\rightarrow+\infty$ :
\[\widehat{\nabla g}_{h_{2}}(X_{t})_{p}\widehat{\nabla g}_{h_{2}}(X_{t})_{q}\underset{k\rightarrow +\infty}{\overset{P}{\longrightarrow}}\nabla\mathbb{E}[(x_{t+s}-\widehat{f}_{h}(X_{t}))^{2}]_{p}\nabla\mathbb{E}[(x_{t+s}-\widehat{f}_{h}(X_{t}))^{2}]_{q}\]
If $\widehat{f}_{h}(X_{t})$ is a consistent estimator of $f$, we get the asymptotic consistency of $\widehat{\nabla g}_{h_{2}}(X_{t})_{p}\widehat{\nabla g}_{h_{2}}(X_{t})_{q}$ toward $\nabla g(X_{t})_{p}\nabla g(X_{t})_{q}$.

Finally, from Theorem \ref{theo1} and if we suppose $h\rightarrow 0$, $h_{2}\rightarrow 0$, $kh^{d}\rightarrow +\infty$, $kh_{2}^{d}\rightarrow +\infty$, $kh^{d+2}\rightarrow +\infty$ and $kh_{2}^{d+2}\rightarrow +\infty$ when $k\rightarrow +\infty$, we have 
\[\widehat{I}_{h,h_{2}}(X_{t})_{pq}\underset{k\rightarrow +\infty}{\overset{P}{\longrightarrow}}I(X_{t})_{pq}\]
where $I(X_{t})_{pq}$ denotes the coefficient on the $p^{th}$ row and $q^{th}$ column of the matrix $I(X_{t})$.

\newpage

\section*{References}
\begin{description}
\item[]{\sc
Arya S., Mount D. M., Netanyahu N. S., Silverman R. and Wu A. Y.} (1998) An optimal algorithm
for approximate nearest neighbor searching, \textit{Journal of the ACM}, Vol.45, pp.891-923.

\item[]{\sc
Bentley J.L.} (1975) Multidimensional binary search trees used for associative searching, \textit{Commun. ACM 18}, Vol.9, pp.509-517. 

\item[]{\sc
Bollerslev T.} (1986) Generalized autoregressive conditional heteroskedasticity, \textit{Journal of Econometrics}, Vol. 31, pp.307-327.

\item[]{\sc
Bollerslev T., Chou R.Y. and Kroner K.F.} (1992) ARCH modeling in finance : a review of the theory and empirical evidence, \textit{Journal of Econometrics}, Vol.52, pp.5-59.


\item[]{\sc
Brock W.A.} (1986) Distinguishing random and deterministic systems, \textit{Journal of Economic Theory}, Vol.40, pp.168-195.

\item[]{\sc
Cao L.} (1997) Practical method for determining the minimum embedding dimension of a scalar time series,  \textit{Physica D : Nonlinear Phenomena}, Vol.110, pp.43-50.



\item[]{\sc
Dämmig M. and Mitschke F.} (1993) Estimation of Lyapunov
exponents from time series: the stochastic case, \textit{Physics Letters A}, Vol.178, pp.385-394.

\item[]{\sc
Dennis B., Desharnais R.A., Cushing J.M., Henson S.M., Costantino R.F.} (2003) Can noise induce chaos?, \textit{Oikos}, Vol.102, N°2, pp.329-339.

\item[]{\sc
Engle R.F.} (1982) Autoregressive conditional heteroscedasticity with estimates of the variance of United Kingdom inflation, \textit{Econometrica}, Vol.50, N° 4, pp.987-1007.


\item[]{\sc
Fan J. and Gibjels I.} (1996) Local polynomial modelling
and its applications, Chapman and Hall, London.

\item[]{\sc
Fraser A. and Swinney H.} (1986) Independent coordinates for strange attractors from mutual information, \textit{Physical Review A}, Vol.33, pp.1134-1140.

\item[]{\sc
Friedman J.H., Bentley J.L. and Finkel R.A.} (1977) An algorithm for finding best matches in logarithmic expected time, \textit{ACM Transactions on Mathematical Software}, Vol.3, pp.209-226. 

\item[]{\sc
Gençay R.} (1996) A statistical framework for testing chaotic dynamics via Lyapunov exponents, \textit{Physica D : Nonlinear Phenomena}, Vol.89, pp.261-266.

\item[]{\sc
Grassberger P. and Procaccia I.} (1983) Measuring the Strangeness of Strange Attractors, \textit{Physica D : Nonlinear Phenomena}, Vol.9, pp.189-208.

\item[]{\sc
Hamilton J.D.} (1994) Time series analysis, Princeton University Press.

\item[]{\sc
Hommes C.H.} (2001) Financial markets as nonlinear adaptive evolutionary systems, \textit{Quantitative Finance}
Vol.1, Issue 1, pp.149-167.

\item[]{\sc
Hommes C.H. and Manzan S.} (2005) Testing for nonlinear structure and chaos in economic time series : a comment, \textit{CeNDEF Working Papers 05-14}, Universiteit van Amsterdam, Center for Nonlinear Dynamics in Economics and Finance.

\item[]{\sc
Hsieh D. A.} (1991) Chaos and Nonlinear Dynamics: Application to Financial Markets, \textit{Journal of Finance},
Vol. 46, No. 5, pp. 1839-1877.

\item[]{\sc
Jeffreys H.} (1946) An invariant form for the prior probability in estimation problems, \textit{Proceedings of the Royal Society of London Series A}, Vol.186, pp.453-461.

\item[]{\sc 
Jolliffe I.T.} (2002) Principal component analysis, Second Edition, Springer-Verlag, New-York.

\item[]{\sc
Kantz H. and Schreiber T.} (2003) Nonlinear time series analysis, Second Edition, Cambridge University Press, Cambridge.

\item[]{\sc
Kawaguchi A. and Yanagawa T.} (2001) Estimating correlation dimension in chaotic time series, \textit{Bulletin of informatics and cybernetics}, Vol.33, pp.63-71.

\item[]{\sc
Kawaguchi A., Yonemoto K. and Yanagawa T.} (2005) Estimating the correlation dimension from a chaotic system with dynamic noise, \textit{Journal of Japan Statistical Society}, Vol. 35, N° 2, pp.287-302.

\item[]{\sc
Kennel M.B., Brown R. and Abarbanel H.D.I.} (1992) Determining embedding dimension for phase-space reconstruction using a geometrical construction, \textit{Physical Review A}, Vol.45, N° 6, pp.3403-3411.

\item[]{\sc
Kyrtsou C., Labys W.C. and Terraza M.} (2004) Noisy chaotic dynamics in commodity markets, \textit{Empirical Economics}, Vol. 29, N° 3, pp. 489-502. 

\item[]{\sc
Kyrtsou C. and Serletis A.} (2006) Univariate tests for nonlinear structure, \textit{Journal of Macroeconomics}, Vol.28, N° 1, pp.154-168.

\item[]{\sc
LeBaron B.} (1992) Some relations between volatility and serial correlations in stock market returns, \textit{Journal of Business}, Vol. 65, N° 2, pp. 199-219.


\item[]{\sc
Lu Z.-Q.} (1999) Multivariate local polynomial fitting for martingale nonlinear regression models, \textit{Annals of the Institute of Statistical Mathematics}, Vol.51, N° 4, pp.691-706. 

\item[]{\sc
Lu Z.-Q. and Smith R.L.} (1997) Estimating local Lyapunov exponents, in \textit{Nonlinear Dynamics and Time Series}, editors Colleen D. Cutler and Daniel T. Kaplan. pp.135-151. Fields Institute Communications Vol. 11, American Mathematical Society, 1997.

\item[]{\sc
Mamon R.S. and Elliott R.J.} (2007) Hidden Markov models in finance, International Series in Operations Research $\&$ Management Science, Springer.

\item[]{\sc
Masry E.} (1996) Mutivariate regression estimation: local polynomial fitting for time series, \textit{Stochastic Processes and their Applications}, Vol.65, N° 1, pp.81-101. 

\item[]{\sc
McCaffrey D.F., Ellner S., Gallant A.R. and Nychka D.W.} (1992) Estimating the Lyapunov exponent of a chaotic system with nonparametric regression, \textit{Journal of the American Statistical Association},
Vol. 87, No. 419, pp.682-695.

\item[]{\sc
Moon Y.-I., Rajagopalan B. and Lall U.} (1995) Estimation of mutual information using kernel density estimators, \textit{Physical Review E}, Vol.52, pp.2318-2321.

\item[]{\sc
Nychka D.W., Ellner S., McCaffrey D.F. and Gallant A.R.} (1992) Finding chaos in noisy systems, \textit{Journal of the Royal Statistical Society Series B}, Vol.54, pp.399-426.


\item[]{\sc
Peters E.E.} (1994) Fractal market analysis : applying chaos theory to investment and economics, Wiley Finance Series, John Wiley $\&$ Sons Inc.

\item[]{\sc
Rosenstein M.T., Collins J.J. and DeLuca C.J.} (1993) A practical method for calculating largest Lyapunov exponents from small data sets, \textit{Physica D : Nonlinear Phenomena}, Vol.65, N° 1-2, pp.117-134.

\item[]{\sc
Sauer T., Yorke J.A., Casdagli M.} (1991) Embedology, \textit{Journal of Statistical Physics}, Vol.65, pp.579-616.

\item[]{\sc Schittenkopf C., Dorffner G. and Dockner E.J.} (2000) On nonlinear, stochastic dynamics in economic
and financial time series, \textit{Studies in Nonlinear Dynamics and Econometrics}, Vol.4, pp.101-121.

\item[]{\sc
Shintani M. and Linton O.} (2003) Is there chaos in the world economy? A nonparametric test using consistent standard errors, \textit{International Economic Review}, Vol.44, N° 1, pp.331-357.

\item[]{\sc
Shintani M. and Linton O.} (2004) Nonparametric neural network estimation of Lyapunov
exponents and a direct test for chaos, \textit{Journal of Econometrics}, Vol.120, pp.1-33.

\item[]{\sc
Takens F.} (1981) Detecting strange attractors in turbulence, in \textit{Dynamical Systems and Turbulence} edited by Rand D.A. and Young L.-S., Springer-Verlag, Berlin.

\item[]{\sc
Tanaka, Aihara and Taki} (1998) Analysis of positive Lyapunov exponents from random time series, \textit{Physica D : Nonlinear Phenomena}, Vol.111, N° 1, pp. 42-50.

\item[]{\sc
Tong} (1983) Threshold models in non-linear time series analysis, \textit{Lecture notes in statistics}, No.21. Springer-Verlag, New York, USA.



\item[]{\sc
Wolf A., Swift J.B., Swinney H.L., Vastano J.A.} (1985) Determining Lyapunov exponents from a time series, \textit{Physica D : Nonlinear Phenomena}, Vol.16, N°3, pp.285-317.

\item[]{\sc
Yao Q., Tong H.} (1994) On prediction and chaos in stochastic systems, \textit{Philosophical Transactions: Physical Sciences and Engineering}, Vol. 348, N° 1688, Chaos and Forecasting, pp.357-369.

\end{description}

\begin{figure}
    \centering
       \includegraphics[scale=0.5]{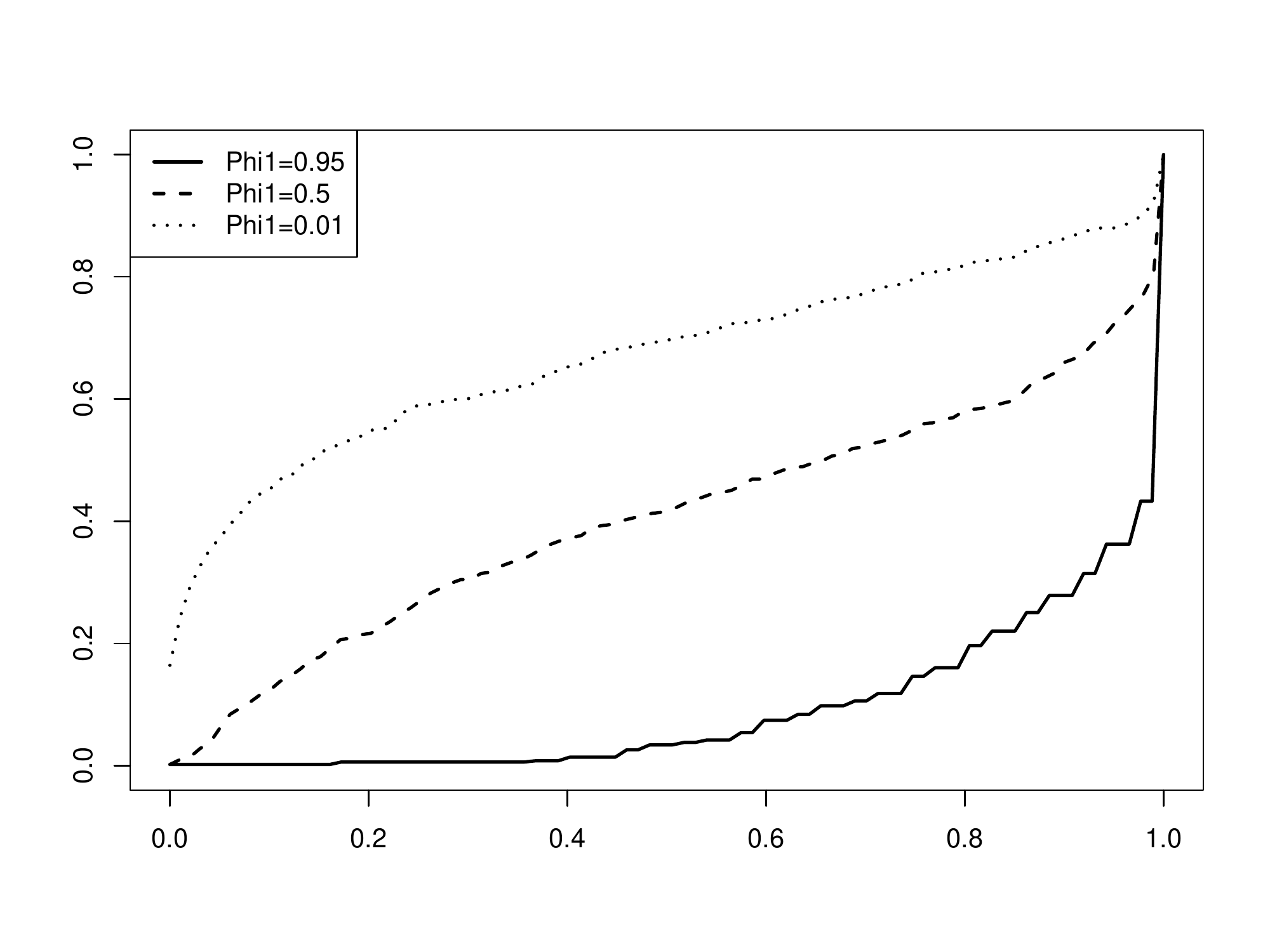}
       \caption{Empirical cumulative distribution functions of the estimated probabilities $\widehat{\mathbb{P}}(S_{h,h_{2}}(T)\leq 0.9)$ for the three AR(1) processes.}
	\label{F1}
\end{figure}

\begin{figure}
    \centering
       \includegraphics[scale=0.5]{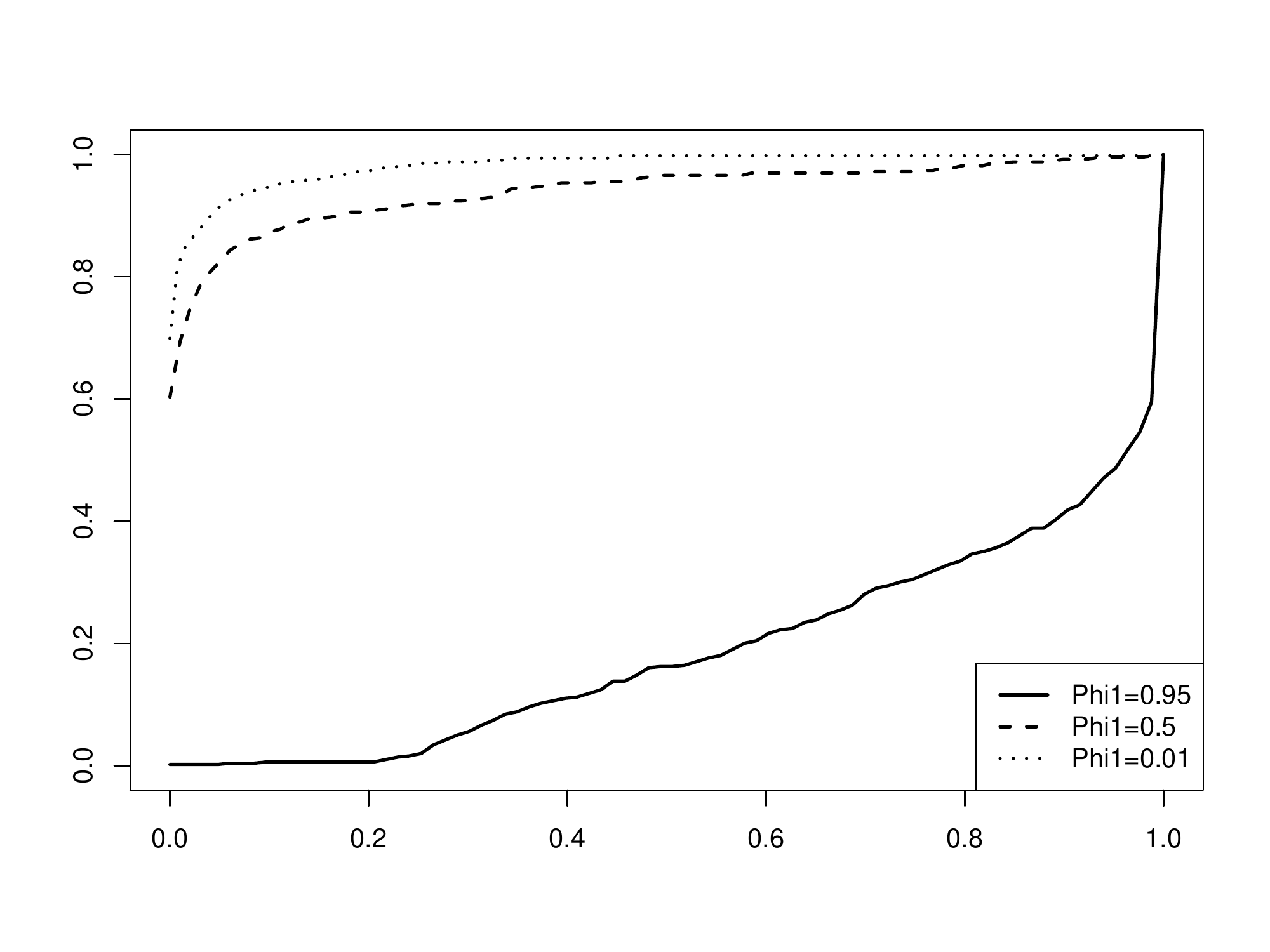}
       \caption{Empirical cumulative distribution functions of the estimated probabilities $\widehat{\mathbb{P}}(S_{h,h_{2}}(T)\leq 0.1)$ for the three AR(1) processes.}
	\label{F2}
\end{figure}

\begin{figure}
    \centering
       \includegraphics[scale=0.5]{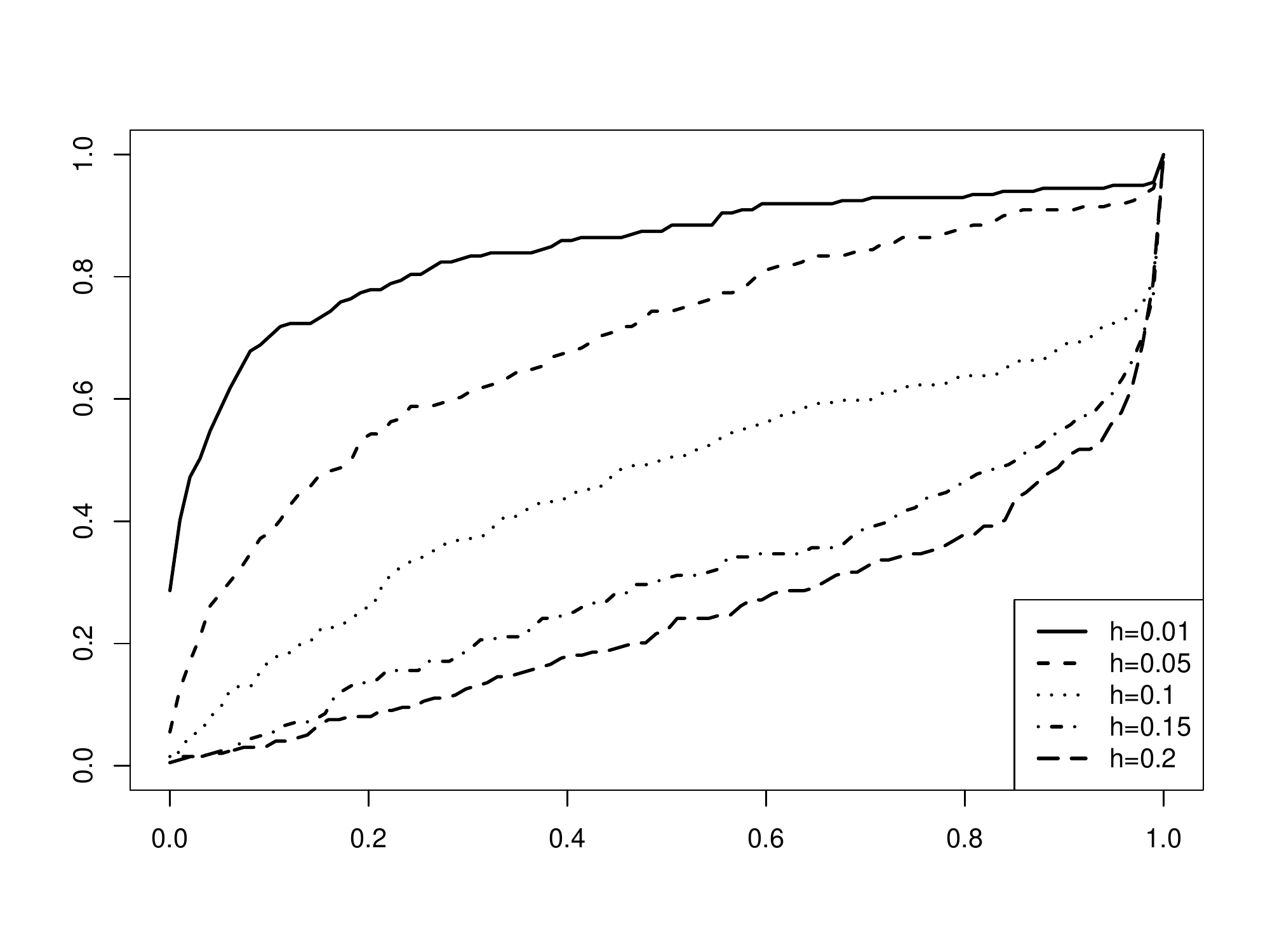}
       \caption{Empirical cumulative distribution functions of the estimated probabilities $\widehat{\mathbb{P}}(S_{h,h_{2}}(T)\leq 0.9)$ in the case $\phi_{1}=0.5$ when $h$ varies.}
	\label{F3}
\end{figure}

\begin{figure}
    \centering
       \includegraphics[scale=0.5]{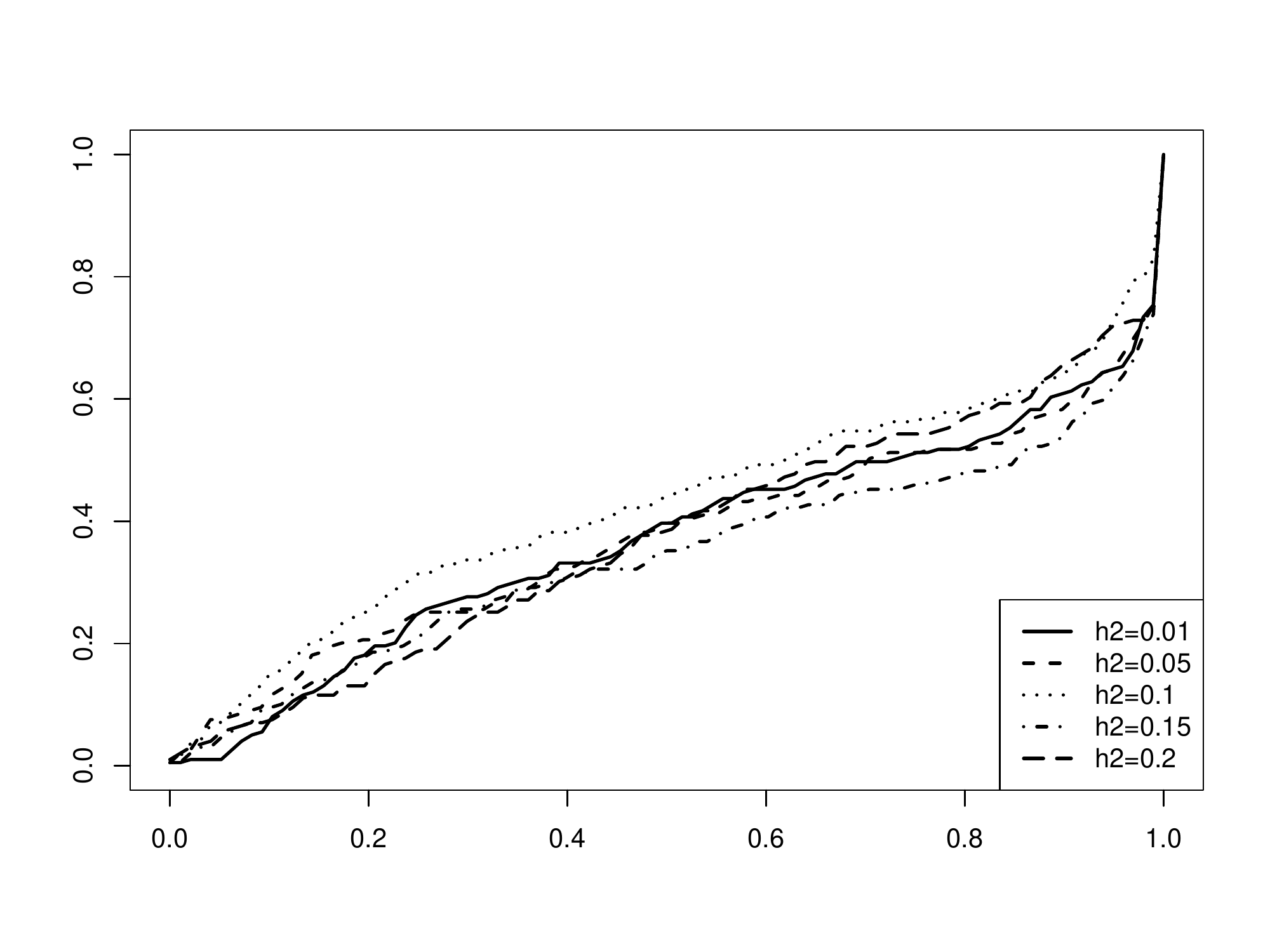}
       \caption{Empirical cumulative distribution functions of the estimated probabilities $\widehat{\mathbb{P}}(S_{h,h_{2}}(T)\leq 0.9)$ in the case $\phi_{1}=0.5$ when $h_{2}$ varies.}
	\label{F4}
\end{figure}

\begin{figure}
    \centering
       \includegraphics[scale=0.5]{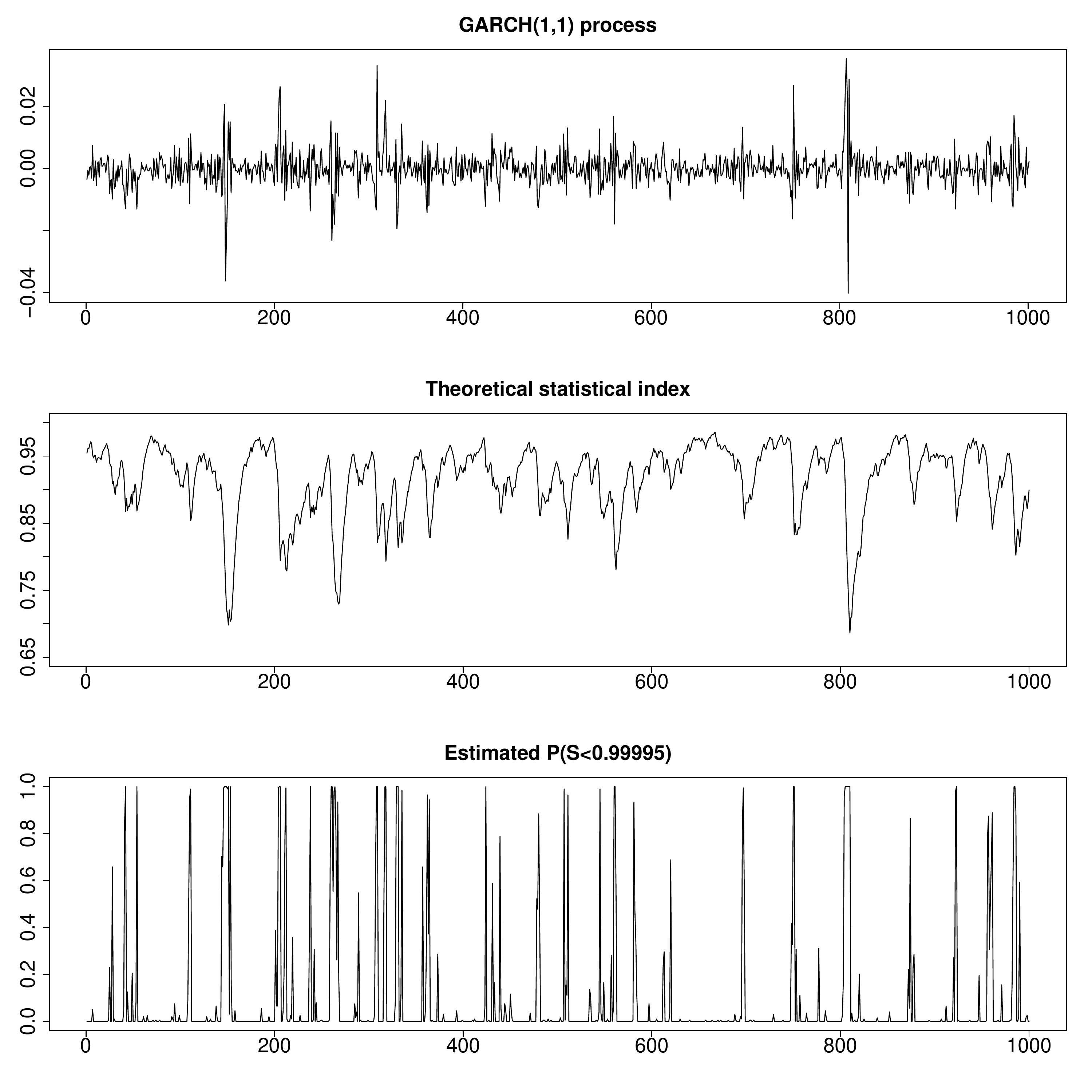}
       \caption{At the top : The simulated GARCH(1,1) process. At the Middle : The corresponding time-varying theoretical statistical index $S(t)$. At the bottom : The corresponding time-varying probability $\widehat{\mathbb{P}}(S_{h,h_{2}}(t)\leq 1-\epsilon)$ with $\epsilon=5\times 10^{-5}$.}
	\label{garch22}
\end{figure}

\
\begin{figure}
    \centering
       \includegraphics[scale=0.5]{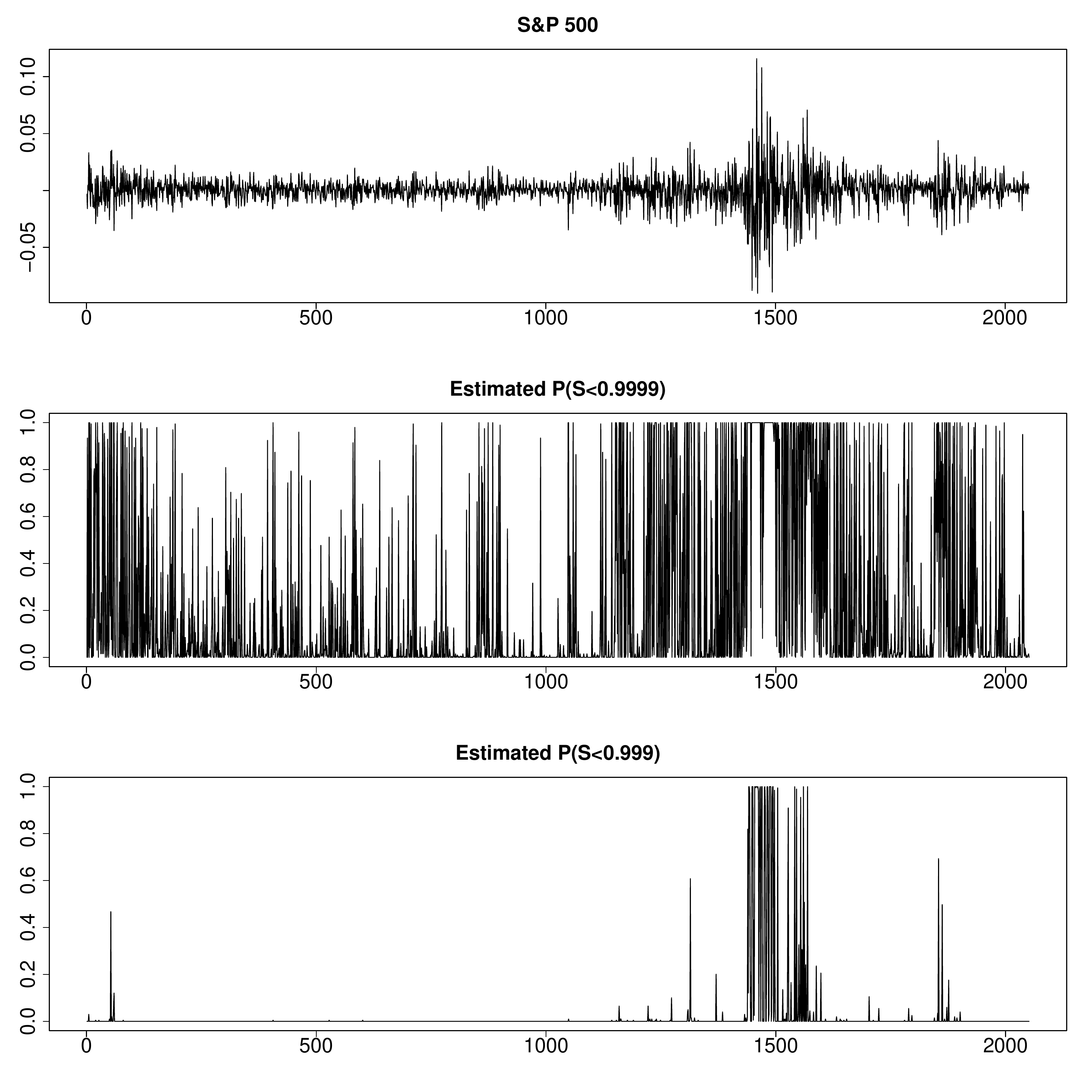}
       \caption{At the top : Daily returns of the $S\&P500$ from the 25/12/2002 to the 18/02/2010. At the Middle : The corresponding time-varying probability $\widehat{\mathbb{P}}(S_{h,h_{2}}(t)\leq 1-1\times 10^{-4})$. At the bottom : The corresponding time-varying probability $\widehat{\mathbb{P}}(S_{h,h_{2}}(t)\leq 1-1\times 10^{-3})$.}
	\label{sp}
\end{figure}

\begin{figure}
    \centering
       \includegraphics[scale=0.5]{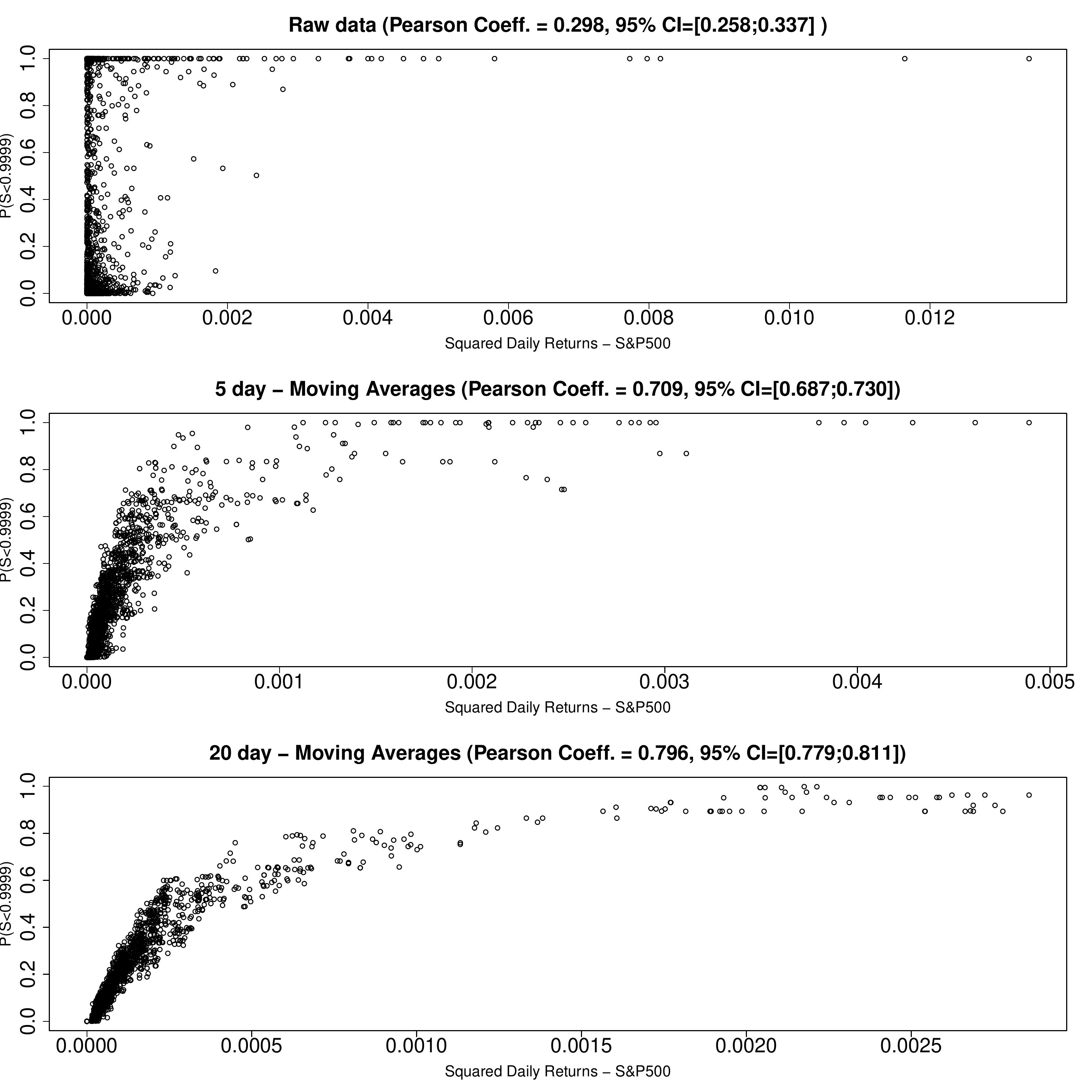}
       \caption{At the top : $\widehat{\mathbb{P}}(S_{h,h_{2}}(t)\leq 1-1\times 10^{-4})$ in function of the squared daily returns of the $S\&P500$ from the 25/12/2002 to the 18/02/2010. At the Middle : 5 day-Moving Average of $\widehat{\mathbb{P}}(S_{h,h_{2}}(t)\leq 1-1\times 10^{-4})$ in function of the 5 day-Moving Average of the squared daily returns of the $S\&P500$ from the 25/12/2002 to the 18/02/2010. At the bottom : 20 day-Moving Average of $\widehat{\mathbb{P}}(S_{h,h_{2}}(t)\leq 1-1\times 10^{-4})$ in function of the 20 day-Moving Average of the squared daily returns of the $S\&P500$ from the 25/12/2002 to the 18/02/2010.}
	\label{PV}
\end{figure}
\end{document}